\documentclass[aps,prd,12point,twocolumn,nofootinbib,showpacs,superscriptaddress]{revtex4-1}

\usepackage{graphicx}
\usepackage{dcolumn}
\usepackage{tabu}
\usepackage{amsmath,amssymb}
\usepackage{float}
\usepackage{natbib}
\usepackage{balance}
\usepackage{braket}
\usepackage{mathrsfs}
\usepackage{bm}
\usepackage{lipsum}
\usepackage{comment}
\usepackage[dvipsnames]{xcolor}
\usepackage[title]{appendix}
\usepackage{txfonts}
\usepackage[breaklinks=true,pdfborder={0 0 0},bookmarksopen]{hyperref}
\hypersetup{
	colorlinks   = true, 
	urlcolor     = Magenta, 
	linkcolor    = blue, 
	citecolor   = Red 
}

\begin{document}

\title{Influence of the dispersion relation on the Unruh effect according to the relativistic Doppler shift method}


\author{F. Hammad} \email{fhammad@ubishops.ca}
\affiliation{Department of Physics and Astronomy, Bishop's University, 2600 College Street, Sherbrooke, QC, J1M~1Z7 Canada}
\affiliation{Physics Department, Champlain 
College-Lennoxville, 2580 College Street, Sherbrooke,  
QC, J1M~0C8 Canada}
\affiliation{D\'epartement de Physique, Universit\'e de Montr\'eal,\\
2900 Boulevard \'Edouard-Montpetit,
Montr\'eal, QC, H3T 1J4
Canada}
\author{A. Landry} \email{alexandre.landry.1@umontreal.ca} 
\affiliation{D\'epartement de Physique, Universit\'e de Montr\'eal,\\
2900 Boulevard \'Edouard-Montpetit,
Montr\'eal, QC, H3T 1J4
Canada} 
\author{D. Dijamco} \email{ddijamco18@ubishops.ca} 
\affiliation{Department of Physics and Astronomy, Bishop's University, 2600 College Street, Sherbrooke, QC, J1M~1Z7 Canada}

\begin{abstract}
We examine the influence of the dispersion relation on the Unruh effect by Lorentz boosting the phase of Minkowski vacuum fluctuations endowed with an arbitrary dispersion relation. We find that, unlike what happens with a linear dispersion relation exhibited by massless fields, thermality is lost for general dispersion relations. We show that thermality emerges with a varying ``apparent" Davies-Unruh temperature depending on the acceleration of the observer and on the degree of departure from linearity of the dispersion relation. The approach has the advantage of being intuitive and able to pinpoint why such a loss of thermality occurs and when such a deviation from thermality becomes significant. We discuss the link of our results with the well-known fundamental difference between the thermalization theorem and the concept of Rindler noise. We examine the possible experimental validation of our results based on a successful setup for testing the classical analogue of the Unruh effect recently described in the literature.
\end{abstract}

\maketitle
\section{Introduction}\label{sec:Intro}
Concepts like modified dispersion relations and minimal length have been introduced into physics through various approaches to quantum gravity \cite{Gross1987,Gross1988,Amati1989,Konishi1990,Maggiore1993}. As soon as they appeared in the literature, they have immediately been put to use as tools to investigate deviations from what is predicted for phenomena relying on Heisenberg's uncertainty principle or those phenomena relying on the usual relativistic dispersion relation (see, e.g., Refs.\,\cite{Lab,GUPWald,Bouda1,Bouda2,Cyclic,Bouda4,Bouda3} and the review papers \cite{Tawfik2014,Tawfik2015}). In addition, it is well known that these two concepts are actually related to each other. A minimal length concept, as implied by the generalized uncertainty principle (GUP) \cite{GUPLiterature1993,GUPLiterature1994,GUPLiterature1995,GUPLiterature2003}, does yield a modified dispersion relation as well \cite{Tawfik2015} and vice-versa \cite{DSR-GUP}. See Ref.\,\cite{DSR-GUP-MDR} for a thorough discussion of the inter-relationships between the various concepts. Therefore, any result obtained under the assumption of modified dispersion relations should also shed some light on cases invoking the generalized uncertainty principle. Our focus in this paper will thus be restricted to the former for which a more intuitive picture of the results can be gained as we shall see. 

One of the very fundamental applications of the concept of modified dispersion relations (or minimal length) is found in the investigation of black hole thermodynamics \cite{GUPBHEntropy,GUP+MDR-BHThermo,Nouicer2007,GEUP2008,BHCorrections2011,Nouicer2012,GUPStat,Progress2019}. In fact, as the Hawking radiation combines quantum field theory with a curved spacetime \cite{Hawking}, one naturally hopes to learn from such an investigation more about semi-classical and quantum gravity by working in deformed settings: either on the classical spacetime side or on the quantum field theory side. These two alternatives offer thus two different approaches. The first possibility allows one to work with a black hole's metric background by using the classical metric itself and its possible deformations \cite{Rainbow} to investigate black hole thermodynamics \cite{WithMDRmetric} and the outgoing Hawking radiation. The second possibility is to take full advantage of the equivalence principle \cite{Cavities}. One can then compute, instead, the spectrum of the deformed quantum fields that would be detected by an accelerated observer in a Minkowski vacuum, {\it i.e.}, using the Unruh effect \cite{Fulling1973,Unruh1973,Davies,Unruh,SovPhys87,QFAccelerated1989}. We shall focus in this paper on the second possibility rather than the first because much less is rigorously known about the effect of modified dispersion relations on spacetime at the {\it quantum} level. In addition, the Unruh effect has been suggested, among other things, to be a potential alternative for investigating non-local field theories \cite{NonLocalFT2018,NonLocalFT} as well as various quantum gravity proposals \cite{QGSignatures}. Moreover, as we shall see, it is conceptually very instructive to mimic the effect of gravity by letting an accelerating observer detect the spectrum of vacuum fluctuations that would obey an arbitrary dispersion relation.

Several authors have already investigated the influence of minimal length and modified dispersion relations on the Unruh effect based on the standard and widely used methods of Bogoliubov transformations and Wightman's two-point functions (for a review, see, e.g., Refs.\,\cite{Takagi86,UnruhReview2008}). The latter emerge within the Unruh-DeWitt point-like detector approach \cite{BirrellDavies}. However, based on these two different methods, different results were reported by various authors. Literature on the effect of minimal length/modified dispersion relations on the Unruh effect can indeed be split into two classes. In one class, one finds reports concluding that the Unruh effect is preserved --- in the sense that thermality emerges --- but that Unruh's temperature acquires a correcting factor. In another class, one learns that the Unruh effect gets destroyed altogether as thermality is lost beyond a certain energy/frequency threshold determined by the minimal length/cutoff frequency imposed. In Ref.\,\cite{UnruhFromGUP2018}, for example, it was found, based on Bogoliubov transformations, that a GUP-inspired modification to the commutation relations still implies an Unruh temperature albeit modified by a factor that is quadratic in the acceleration $a$ of the observer. In Ref.\,\cite{UnruhFromGUP2019}, on the other hand, Wightman's function is combined with the generalized proper-time proposal of Ref.\,\cite{MaximalA} to arrive at a different dependence of Unruh's temperature on $a$ (from which a constraint on the concept of maximal acceleration was suggested). Very important also, is that in these references the mass of the detected Rindler particles does not seem to affect the results. In Ref.\,\cite{GUP-EUP2019}, yet another expression for Unruh temperature in terms of acceleration $a$ is obtained based on the so-called extended uncertainty principle (EUP), which introduces the notion of minimal momentum through a modified time-energy uncertainty relation. 

Instead of making use of modified dispersion relations, a modified Wightman's function was used in Ref.\,\cite{InvariantPlanck} to compute the power spectrum which was found to be thermal only up to a certain frequency scale that depends on the minimal length introduced inside the deformed Wightman's function (see also Ref.\,\cite{PlanckScale}). On the other hand, using the Unruh-DeWitt detector method combined with a specific dispersion relation for the massless photons, it was found in Ref.\,\cite{MDRUnruh} that the power spectrum of the detected modes of the Rindler particles deviates from the Planck spectrum by a frequency-dependent factor such that only a limited range of the original vacuum frequencies yields a positive spectrum (see also, Ref.\,\cite{LorentzBoost}). In Ref.\,\cite{Minimal-LUnruh}, the Unruh effect was examined by using Wightman's function extracted from a deformed propagator as implied by minimal length. It was found there that the Unruh effect disappears as thermality is also lost, to be recovered only for accelerations $a$ of the observer below a threshold fixed by the minimal length. Similarly, using the particle detector approach, it was found in Refs.\,\cite{KappaUnruh2007,KappaUnruh2012} that the Unruh effect disappears exponentially as the proper time of the detector (observer) exceeds a certain threshold fixed by the parameter $\kappa$ within a $\kappa$-Minkowski spacetime \cite{Kappa1992,NonCommut1995} in which the commutation relations of the quantum field are deformed as well. In Ref.\,\cite{Superluminal1}, the effect on Unruh radiation of a superluminal dispersion relation with a very specific form has been investigated using Wightman's function. It was concluded that the Unruh effect remains a low energy phenomenon as the correction to thermality is inversely proportional to the square of the cutoff scale in the modified dispersion relation.  

Now, it turns out that the approach based on the Unruh-DeWitt point-like detector can also be viewed as a relativistic Doppler shift calculation \cite{SovPhys87,PlaneWaves00,PlaneWaves0,PlaneWaves1,PlaneWaves2}. Although such a point of view is extremely intuitive, it has not been much mentioned in the literature regarding the Unruh effect, and never discussed regarding the case of modified dispersion relations. Nevertheless, one of the advantages of the approach is to provide an intuitive picture for the Unruh effect by showing that the latter is deeply rooted in classical physics as well (see also, Refs.\,\cite{UnruhRadiation?,ClassicalRoots1,UnruhRaditation}). Another advantage of the approach is to naturally provide a description of the effect in terms of spontaneous and induced emissions of particles by the detector leading to a thermal spectrum thanks to Einstein's detailed balance equation for systems in thermal equilibrium \cite{SpontaneousInduced}. We thus propose in this paper to examine the influence of modified dispersion relations on the Unruh effect by analyzing the spectrum perceived by an accelerated observer as the former is shaped by the relativistic Doppler shift caused by the accelerated motion of the latter. The approach has one more advantage of being very general as it easily works for an arbitrary dispersion relation of the field under consideration. Furthermore, it provides a precise and a very clear intuitive picture of why there might be a loss of thermality in the detected spectrum. The superiority of the approach over the Bogoliubov transformations method when dealing with arbitrary dispersion relations or the minimal length via the GUP consists also in the fact that these imply a modification of the equations of motion of the field and its propagator (see, e.g., Refs.\,\cite{MDREFT,GEUP2013,Lifshitz}), rendering thus Bogoliubov transformations very difficult to extract.

The appearance of the Unruh effect as a result of the relativistic Doppler shift of the vacuum fluctuations has been very pedagogically exposed, and in much greater detail, in Ref.\,\cite{Simplified2004}. In the latter reference, the authors started by applying the approach to classical plane waves and then showed how the procedure easily adapts to quantized fields. For massless scalar fields, the method gives back the usual Planck spectrum and Unruh temperature. For massless spin-$\frac{1}{2}$ fields, the method gives back the Fermi-Dirac spectrum \cite{Takagi86,Simplified2004}. Thus, because the approach yields the same results as the Bogoliubov transformations and Wightman's function approaches for the case of massless fields, it never seemed necessary to apply the approach to the case of massive fields and, more generally, to the case of fields obeying arbitrary dispersion relations. It is, however, well known that one does not recover a thermal spectrum with massive fields based on the Unruh-DeWitt detector method. This is in contrast to what the Bogolubov transformations-based approach seems to suggest. The fundamental reason behind this fact, as elaborately explained in Ref.\,\cite{Takagi86}, is deeply rooted in the ``thermalization theorem'', extracted from the Bogoliubov transformations, as opposed to the 
``Rindler noise'' associated with the point-like detector approach. As we shall see, the relativistic Doppler shift calculation readily allows one to clearly see why the Unruh-DeWitt detector method does not yield a thermal spectrum in the massive case. Moreover, the procedure is easier to apply ---\! both formally and conceptually\! --- to arbitrary dispersion relations thanks to its very intuitive nature.

The remainder of this paper is structured as follows. In Sec.~\ref{sec:II}, we expose the intuitive feature of the relativistic Doppler shift approach by applying it to classical plane waves obeying an arbitrary dispersion relation. We show how and why the Planck spectrum is lost when the dispersion relation of the waves departs from the massless case. In Sec.~\ref{sec:III}, we discuss briefly how to apply the approach to quantized fields obeying an arbitrary dispersion relation. In Sec.\,\ref{Implications}, we examine in detail the experimental implications of our results and their possible validation using the recently successful experimental setup described in Ref.\,\cite{Analog2018}. The more involved calculations in this paper are gathered in the two appendices \ref{A} and \ref{B}. We conclude this paper with a brief summary.
\section{Classical fields}\label{sec:II}
The first goal of this section is to acquire some intuition for the relativistic Doppler shift method by applying the latter to the case of classical plane waves. The other goal is to derive the useful equations that can be adopted to quantum fields and that we shall adopt in Sec.\,\ref{Implications} for an eventual experimental test. We shall therefore examine here the spectrum detected by an accelerating observer by working out the phase shift caused by the motion of the latter on classical plane waves obeying an arbitrary dispersion relation. 

First, recall that for a particle of mass $m$, of energy $E$, and of three-momentum $\bf p$, the usual relativistic dispersion relation reads, $E^2={\bf p}^2c^2+m^2c^4$. The simplest example of a modified dispersion relation for massless particles, often encountered in the literature, has the form $E^2={\bf p}^2c^2+\beta f({\bf p})$, where the parameter $\beta$ determines the energy scale at which such a modification becomes relevant (see Ref.\,\cite{MDRVersions} for examples of non-relativistic versions). For the sake of generality, we shall consider here an arbitrary dispersion relation of the form $E=f(\bf P)$, inside which a nonzero mass might be included. In order to be able to work with waves, however, we have to consider instead an arbitrary relation between the angular frequency $\omega$ and the wave vector $\bf k$ which might be extracted from the momentum $\bf p$ and energy $E$ thanks to general expressions of the form $\omega=f_0(E,{\bf p})$ and ${\bf k}=f_1(E,{\bf p})$, respectively \cite{MDRVersions}\footnote{For ease of notation, we shall set in this section the Planck constant $\hbar$ as well as the speed of light $c$ to unity.}. Thus, a general modified dispersion relation might be taken to be of the form $\omega=f(\bf k)$ for a regular and smooth function $f$. For the massless non-deformed case, such a dispersion relation reduces to the linear relation $\omega=|\bf k|$. As the generality of our approach is already guaranteed by the arbitrary function $f$, we shall assume isotropy and consider only the one-dimensional case for which the results become more transparent and free from extra unnecessary transverse terms.

Let us therefore consider a one-dimensional classical plane wave obeying the dispersion relation, $\omega=\omega(k)$, for an arbitrary dependence of the angular frequency $\omega$ on the wave number $k$. As we require that the modified dispersion relation reduces to a standard one at low energies and still guarantee a positive $\omega$ for high energies, we assume that $\omega(k)\geq k$ for classical and quantum fields. Therefore, at any given point in Minkowski spacetime, the phase of a plane wave obeying such a dispersion relation and moving either in the negative or positive direction, respectively, is of the form, $e^{i\phi_{\pm}(t,x)}=e^{i[\omega(k) t\pm kx]}$. Let us then examine how this phase becomes affected by the motion of an accelerating observer and what phase would the latter observe by following the same steps exposed in Ref.\,\cite{Simplified2004}. 

First, the Minkowski time $t$ and position $x$ are related to the proper time $\tau$ of an observer moving, say in the positive $x$-direction, with constant acceleration $a$, by the following Rindler coordinate transformation:
\begin{equation}\label{x,t}
t(\tau)=\frac{\sinh{a\tau}}{a},\qquad x(\tau)=\frac{\cosh{a\tau}}{a}.
\end{equation}
Therefore, substituting these expressions for $x$ and $t$ inside the above expression of the phase in Minkowski spacetime, we find the effective phase detected by the observer to be of the form
\begin{equation}\label{ObserverPhase}
e^{i\phi_{\pm}(\tau)}=\exp\left(i\left[\frac{\omega(k)\pm k}{2a}e^{a\tau}- \frac{\omega(k)\mp k}{2a}e^{-a\tau}\right]\right).
\end{equation}
Before we examine the resulting detected spectrum as it arises from this expression of the effective phase, a couple of important remarks concerning (i) our use of plane waves and (ii) our assignment of the hyperbolic motion (\ref{x,t}) to the accelerating observer are in order here. 

The first remark concerns our use of plane waves with a modified dispersion relation. Since our plane waves exhibit a four-momentum that obeys modified dispersion relations, these plane waves are taken to be solutions of modified wave equations in the case of classical waves and solutions of modified field equations of motion in the case of free quantum fields \cite{NonLocalFT2018,MDREFT,GEUP2013,Lifshitz,DSRInx2004,Lehnert,2019a,2019b,2020}. The modified wave equations being linear, the plane wave solutions are always guaranteed to exist provided only that one requires that the contraction $p_\mu x^\mu$ between position and momentum remains linear, {\it i.e.}, $p_\mu x^\mu=p_0x^0+p_ix^i$ \cite{Rainbow,DSRInx2004}. It is, actually, such a requirement that allows one to have linear modified Lorentz transformations in position space even though such transformations are nonlinear in momentum space \cite{DSRInx2004,DSRLiterature2005,DSRInx2007}. In addition, it is such a requirement that leads to the elegant interpretation of the modified dispersion relations in terms of an energy-dependent spacetime metric as ``seen'' by a quantum particle thanks to the existence of a modified quadratic invariant \cite{DSRInx2004,DSRLiterature2005}, {\it i.e.}, the so-called gravity's rainbow \cite{Rainbow}. 

On the other hand, minimal length, as implied by a non-commutativity of the position and momenta operators, still allows one to expand a quantum field as usual in terms of plane waves weighted by creation and annihilation operators \cite{Minimal}. If, however, one decides to replace plane waves by the so-called ``maximal localization states'' of the GUP-modified commutation relations \cite{GUPLiterature1995,MaxGUP1,MaxGUP2}, then one would lose any trace of the Planck spectrum in this approach because of the loss of a meaningful separation between the position $x$ of the observer and the position expectation value $\braket{\hat x}$ in the field expansion. Indeed, one would then associate to the field phases of the form \cite{QFTFromGUP}: $\exp\left\{{i\,\left[\omega(k)t\,\pm\,\frac{\braket{\hat x}}{\sqrt{\beta}}\tan^{-1}(\sqrt{\beta}\, k)\right]}\right\}$. This expression of the phase does not allow the emergence of the Planck spectrum because of the lack in it of a symmetry between the position and time coordinates. Such a symmetry is indeed required to give rise to the crucial term $e
^{-\pi\Omega/2a}$, as we shall see in detail shortly.  

The second remark concerns our assignment of a hyperbolic motion of the form (\ref{x,t}) to the observer. In fact, one should recall that such a description of the motion in terms of the proper time $\tau$ of the observer is obtained based on the usual Lorentz transformations. Being here interested instead in waves and fields that obey modified dispersion relations, which automatically violate Lorentz invariance, one might wonder what allows us to keep using Lorentz boosts. The reason behind assigning such a hyperbolic motion to our observer is that we assume the latter's trajectory to be independent of the fields and particles he/she is supposed to detect. The observer ---\! like a detector\! --- is taken here to be a macroscopic object that is not altered by the quantum fluctuations of the background spacetime inside which it propagates. This is unlike the quantum particles and fields whose modified dispersion relations are precisely due to their interaction with the background spacetime \cite{DSRInx2004,Rainbow} (see also Ref.\,\cite{MDRVersions} for an interpretation in terms of an induced particle species-dependent pseudo-Finslerian geometry). It must be noted in this regard that such modified Lorentz transformations cannot actually be consistently applied to bound systems of particles as the latter may exceed the Planck mass leading to the so-called ``soccer-ball problem'' \cite{SoccerBall}. Therefore, while the waves/ fields obey modified Lorentz transformations, the observer's position and time coordinates $(x,t)$ displayed in Eq.\,(\ref{x,t}) ---\! at which the 
``onboard sensor'' is located, independently of the waves/fields hitting it\! --- still obey the usual Lorentz transformations. This is the natural approach which is consistent with a Lorentz transformation that applies to a passive observer moving along the usual macroscopic trajectory while being hit by random waves/fields, regardless of when or where the latter have been created and where they are coming from.

If, instead, one is interested in finding the spectrum of the waves (or the vacuum fluctuations) as seen by the observer under the influence of the waves/fields themselves ---\! which are thus used as {\it probes} for the spacetime location of the observer\! --- then one has to use the deformed Lorentz transformations, not the linear ones. In other words, one takes in this case Lorentz transformations to be active transformations in the sense that the detected waves/fields are monitored by the observer from the moment of their creation to the moment of their detection. While such an approach does not square well with the picture of an observer moving independently of the background and randomly hit by these waves/fields, it is, nevertheless, very instructive to examine such a possibility, as we do it in detail in Appendix \ref{B}. Although it is much easier to anticipate for such a case that the detected spectrum would never be Planckian ---\! due to the combination of highly nonlinear modified Lorentz transformations with modified dispersion relations back-reacting on the transformations\! --- the approach based on a macroscopic observer obeying the usual Lorentz transformations provides results that are physically much richer as we shall see now.

To get the shape of the spectrum as detected by the accelerating observer, we Fourier transform the $\tau$-dependent phase (\ref{ObserverPhase}) using an arbitrarily chosen angular frequency $\Omega$ among the continuous spectrum of frequencies accessible to the observer. As the transform consists in evaluating the integral, $g_{\pm}(\Omega)=\int_{-\infty}^{+\infty} e^{i\Omega\tau}e^{i\phi_{\pm}(\tau)}{\rm d}\tau$, we are going to perform the change of variable $e^{a\tau}=y$. Such a Fourier transform then simplifies greatly and reduces to 
\begin{equation}\label{Transform}
g_{\pm}(\Omega)=\frac{1}{a}\int_{0}^{\infty} y^{\pm\nu-1}e^{\pm i\left(\xi y-\frac{\eta}{y}\right)}{\rm d}y.
\end{equation}
Here, we have distinguished the spectrum $g_{+}(\Omega)$ of the left-moving modes from the spectrum $g_{-}(\Omega)$ of the right-moving ones. We have also set, for convenience,
\begin{equation}\label{XiEta}
\nu=i\frac{\Omega}{a},\qquad \xi=\frac{\omega(k)+k}{2a},\qquad\eta=\frac{\omega(k)-k}{2a}.
\end{equation}

To evaluate integral (\ref{Transform}), it is actually more practical to split the exponential function into the complex sum of a cosine and a sine function (see Eq.~(\ref{CosineSine}) of Appendix \ref{A}). In fact, the integral then becomes easier to evaluate by using the tables of integrals given in Ref.\,\cite{FormulasBook}. Thus, the expressions of the Fourier amplitudes $g_{\pm}(\Omega)$ in each wave are given in terms of the modified Bessel function $K_\nu(z)$ \cite{FormulasBook} in the following form,
\begin{equation}\label{g+-}
g_{\pm}(\Omega)=\frac{2\,e^{i\frac{\pi\nu}{2}}}{a}\left(\frac{\eta}{\xi}\right)^{\pm\frac{\nu}{2}} K_{\pm\nu}\left(2\sqrt{\xi\eta}\right).
\end{equation}
The amplitudes $g_{\pm}(\Omega)$ have thus been found in terms of the modified Bessel function $K_\nu(z)$, but could also be expressed in terms of the first Hankel's function $H_\nu^{(1)}(z)$ by using the well-known link (\ref{BesselHankel}) between these two functions \cite{FormulasBook}. The expression in terms of the modified Bessel function $K_\nu(z)$ will allow us to easily find an approximation for $g_{\pm}(\Omega)$ when $\xi\eta$ is very large, whereas the expression in terms of Hankel's function  $H^{(1)}_{\nu}(iz)$ will allow us to find an approximation for $g_{\pm}(\Omega)$ when $\xi\eta$ is very small.

Now, we already see from this result that, in contrast to what one finds when the simple dispersion relation $\omega=k$ holds (valid for massless particles), the spectrum that emerges for an arbitrary dispersion relation cannot be Planckian anymore. In fact, for the latter to show up the final expression of the spectrum (\ref{g+-}) should display the term $\Gamma(\nu)$ which leads to the famous denominator $(e^{2\pi\Omega/a}-1)$ which is characteristic of the Planck spectrum\footnote{When the dispersion relation reduces to the linear one $\omega=k$, expression (\ref{g+-}) is, of course, not valid anymore for in this case $\eta=0$ and expression (\ref{g+-}) becomes ill-defined. For this special case, integral (\ref{Transform}) reduces to $g_{\pm}(\Omega)=\int_0^{\infty}y^{\pm\nu-1}e^{\pm i\xi y}{\rm d}y$. This is, in fact, the integral that leads to the usual Planckian spectrum as it is proportional to $\Gamma(\nu)$ which gives rise, thanks to the property (\ref{Gamma}), to $1/\sinh(i\pi\nu)$ from which, in turn, one obtains the crucial term $1/(e^{2\pi\Omega/a}-1)$ by using the definition (\ref{XiEta}) of $\nu$.}. 

In order to search for any hidden Planckian spectrum inside our result (\ref{g+-}), we are going to dissect the latter by examining the two extreme cases of $\xi\eta\gg1$ and $\xi\eta\ll1$. These would represent, respectively, cases of small and large accelerations $a$ compared to the angular frequency $\omega$. However, we should keep in mind that the case $\xi\eta\gg1$ could also arise for any finite acceleration $a$ of the observer as long as the dispersion relation of the plane wave departs greatly from the linear dispersion relation $\omega=k$ of massless particles. In other words, the case $\xi\eta\gg1$ could also arise for arbitrary accelerations $a$ but with very large deformations of the dispersion relation $\omega=k$. Similarly, the case $\xi\eta\ll1$ could also arise for any finite acceleration $a$ with small deformations of the dispersion relation, {\it i.e.}, as long as the dispersion relation of the wave becomes very close ---\! but not identical\! --- to the linear dispersion relation $\omega=k$.

We shall discuss now the two cases separately by using the general infinite series expansion of Hankel's function $H^{(1)}_{\nu}(z)$ valid for any complex argument $z$.
\subsection{Small accelerations and/or large deformations}\label{sec:II.Smalla}

For small accelerations $a$ compared to the angular frequency $\omega$ of the wave and/or for large departures from the linear dispersion relation $\omega=k$, we have $\xi\eta\gg1$. Using the large-argument expansion of the modified Bessel function (\ref{K>>1}), we find the following approximations at the lowest order in $1/\xi\eta$,
\begin{equation}\label{G>>1}
g_{\pm}(\Omega)\approx\frac{e^{-\frac{\pi\Omega}{2a}}}{a}\sqrt{\frac{\pi}{\sqrt{\xi\eta}}}\left(\frac{\eta}{\xi}\right)^{\pm i\frac{\pi\Omega}{2a}}e^{-2\sqrt{\xi\eta}}.
\end{equation}
We clearly see from this expression that there is no way for the Planck spectrum, {\it i.e.}, for the factor $\Gamma(\nu)$, to be recovered by squaring the amplitudes $g_{\pm}(\Omega)$ and their complex conjugates. In fact, taking the squared magnitude of $g_{\pm}(\Omega)$ we find the following unique result for both amplitudes $g_{\pm}(\Omega)$,
\begin{equation}\label{|G|>>1}
|g_{\pm}(\Omega)|^2\approx\frac{\pi}{a^2\sqrt{\xi\eta}}e^{-\frac{\pi\Omega }{a}}e^{-4\sqrt{\xi\eta}}.
\end{equation}
These amplitudes are exponentially decreasing and the usual denominator $(e^{2\pi\Omega/a}-1)$, characteristic of the Planck spectrum, is clearly missing for any $\eta$.

The natural physical interpretation of this result is as follows. For small accelerations of the observer and/or large deformations of the dispersion relation, the relativistic Doppler shift that affects the original frequencies $\omega$ of the plane waves is not sufficient to give the latter the shape of the Planck distribution. In other words, in contrast to the linear case $\omega=k$, the relativistic Doppler shift of the original spectrum of the plane waves becomes in this case overwhelmed and completely veiled behind the nonlinearity of the deformed dispersion relation. 
\subsection{Large accelerations and/or small deformations}\label{sec:II.Largea}

For large accelerations $a$ compared to the specific angular frequency $\omega$ and/or for small deformations of the dispersion relation, we have $\xi\eta\ll1$. First, using the relation (\ref{BesselHankel}) between the modified Bessel function $K_\nu(z)$ and Hankel's function $H_\nu^{(1)}(iz)$ we easily re-express the amplitudes (\ref{g+-}) in terms of the latter. Then, using the small-argument expansion (\ref{Hankel<<1}) of Hankel's function, we arrive at the following approximation at the leading order in $\xi\eta$,
\begin{align}\label{G<<1}
g_{\pm}(\Omega)&=\frac{i\pi }{a}\left(\frac{\eta}{\xi}\right)^{\pm\frac{\nu}{2}}e^{i\frac{\pi}{2}(\nu\pm\nu)}H_{\pm\nu}^{(1)}(2i\sqrt{\xi\eta})\nonumber\\
&\approx\frac{e^{\frac{i\pi\nu}{2}}}{a}\left[\Gamma(\mp\nu)\,\eta^{\pm\nu}+\Gamma(\pm\nu)\,\xi^{\mp\nu}\right].
\end{align}
In the first line, the factor $e^{\frac{i\pi}{2}(\nu\pm\nu)}$ cancels of course from the amplitude $g_{-}(\Omega)$. By computing the square of the magnitudes from this expression, we find the following unique result for both amplitudes $g_\pm(\Omega)$,
\begin{align}\label{|G|<<1}
|g_{\pm}(\Omega)|^2
&\approx\frac{2e^{-\frac{\pi\Omega}{ a}}}{a^2}\left|\Gamma\left(\frac{i\Omega }{a}\right)\right|^{2}\left(1+\cos\left[2\theta-\frac{\Omega }{a}\ln(\xi\eta)\right]\right).
\end{align}
To arrive at the expression in the square brackets we have used the fact that for the purely imaginary parameter $\nu$, we have the identity ${\rm Re}\left([\Gamma(\nu)]^2(\xi\eta)^{-\nu}\right)$ $=|\Gamma(\nu)|^2\cos\left[2\theta-\frac{\Omega }{a}\ln(\xi\eta)\right]$, where $\theta=\arg\Gamma(\nu)$.
Therefore, for large accelerations of the observer and/or large deformations of the dispersion relation the spectrum reads,
\begin{align}\label{|G+-|<<1}
|g_{\pm}(\Omega)|^2&\approx\frac{8\pi }{a\Omega\left(e^{\frac{2\pi\Omega }{a}}-1\right)}\cos^2\left[\theta-\frac{\Omega }{2a}\ln(\xi\eta)\right].
\end{align}
The Planck spectrum is thus recovered with a specific correcting factor. If we were to interpret this last expression in terms of the Unruh effect, we would conclude that the observer should detect a slightly deformed thermal spectrum with an Unruh temperature given by $T=a/2\pi$\footnote{For convenience, we also set in the rest of this paper the Boltzmann constant $k_B$ equal to unity.}. 

This can be interpreted physically as follows. For large accelerations of the observer and/or small deformations of the dispersion relation the relativistic Doppler shift of each original angular frequency $\omega$ of the plane waves is large enough that deviations of the plane waves from the linear dispersion relation $\omega=k$ have no noticeable effect on the global shape of the resulting spectrum. The latter then takes the same form as the one obtained for the case of the linear dispersion relation except for a minor correcting factor which is closer to unity towards the higher-frequency side of the spectrum. Yet, it is clear that the multiplicative factor $\cos^2[\theta-\Omega\ln(\xi\eta)/2a]$ in Eq.\,(\ref{|G+-|<<1}) is frequency-dependent and thus does deform the global shape of the detected spectrum towards the lower-frequency side of the latter. Nevertheless, we can still extract an ``apparent'' Unruh temperature with a specific frequency-dependent correction as follows. 

For large accelerations and low frequencies making $\Omega\ll a$, the argument $\theta$ of the function $\Gamma(\nu)$ can be approximated to the third order in $\Omega/a$ by
\begin{equation}
\theta\approx\frac{\pi}{2}-\frac{\Omega\gamma}{a}-\left(\frac{\Omega\gamma}{a}\right)^3\left(\frac{1}{6}+\frac{\pi^2}{12\gamma^2}\right),
\end{equation}
where $\gamma$ is the Euler-Mascheroni constant \cite{FormulasBook}. Therefore, the multiplicative factor $\cos^2[\theta-\Omega\ln(\xi\eta)/2a]$ in formula (\ref{|G+-|<<1}) can be approximated up to the third order in $\Omega/a$ as well. Then, factoring out from such an approximation the $\Omega$-independent term, the deviation from the usual Planck spectrum takes the following form,

\begin{equation}\label{Amplitude}
|g_{\pm}(\Omega)|^2\approx\frac{2\pi\Omega }{a^3\left(e^{\frac{\Omega }{T}}-1\right)}\left[2\gamma+\ln\left(\frac{\omega^2-k^2}{4a^2}\right)\right]^2,
\end{equation}
where we have introduced in the denominator the following apparent Unruh temperature,
\begin{equation}\label{Temperature}
T\!\approx\!\frac{a}{2\pi}\!\left[1\!+\!\left(\frac{\Omega}{a}\right)^{\!2}\!\frac{4\pi^2\gamma\!-\!12\gamma^2\ln\frac{\omega^2-k^2}{4a^2}\!-\!6\gamma\ln^2\frac{\omega^2-k^2}{4a^2}\!-\!\ln^3\frac{\omega^2-k^2}{4a^2}}{24\gamma+12\ln\tfrac{\omega^2-k^2}{4a^2}}\!\right]\!.
\end{equation}
The power spectrum is clearly frequency-dependent and expression (\ref{Temperature}) could be identified with a genuine temperature only for the low frequencies $\Omega$ of the spectrum. It does not only depend on the probed frequency $\Omega$, but it depends even on the frequency $\omega(k)$ of the particular wave that has been Doppler shifted. As we shall see in the next two sections, these results and interpretations still hold when the procedure is correctly applied to quantized fields and to waves on a water surface.
\section{Quantized fields}\label{sec:III}
Let us now discuss the relativistic Doppler shift calculation approach for an observer accelerating in a Minkowski vacuum, {\it i.e.}, by taking into account the fluctuations of a quantum field, which we shall take here to be a scalar and neutral quantum field for simplicity. 

As mentioned in the Introduction, the approach based on the linear dispersion relation $\omega=k$ has also been successfully applied to the case of a quantum fermion field in Ref.\,\cite{Simplified2004}. The conclusion drawn in that reference was that the Fermi-Dirac distribution arises naturally as a relativistic Doppler shift effect provided that one takes into account the behavior of spinors under Lorentz boosts. Indeed, the only difference from the scalar field case is the additional Fermi-Walker transport of the fermion field that one has to perform to take into account the effect of the different observer's instantaneous velocities on the phase of the detected spinor field during motion. The crucial multiplicative factor $1/\cosh(i\pi\nu)$ that gives rise to the Fermi-Dirac distribution, rather than the factor $1/\sinh(i\pi\nu)$, then emerges naturally \cite{Simplified2004}. Given that such an extra factor is added as a multiplicative factor that is independent of the dispersion relation, our conclusions in this section concerning the scalar field will remain valid for the case of the fermion field. In other words, the conditions we find for the appearance of the Planck distribution in the case of the scalar field will also be valid for the appearance of the Fermi-Dirac distribution in the case of a fermion field.

Now, the intuition and the physical picture we gained in the previous section concerning the distinction between small/large accelerations and deformations of the linear dispersion relation by dealing with classical waves will constitute a great input here. In addition, however, a richer physical picture and interpretation are now involved as the procedure invokes not just a varying relativistic Doppler shift, but also creation of quanta caused by the accelerated motion of the observer. 

Indeed, in order to be consistent with what we just did for classical waves, we need to Fourier transform only the phases that accompany the operators $a_k$ and $a^\dagger_k$ in the mode expansion of quantum fields, rather than Fourier transforming the whole quantum field operator itself. As a consequence, the random phases of the Minkowski vacuum fluctuations become automatically Doppler-shifted from the point of view of the accelerating observer. One then only needs to compute the squared magnitudes of the resulting Fourier transformed phases. With this way of proceeding, one is guaranteed to recover the results of Sec.\,\ref{sec:II} where we dealt with classical waves. It is indeed clear that this way of applying the approach will just take one through all the steps taken in Sec.\,\ref{sec:II}, starting from Eq.\,(\ref{ObserverPhase}) and all the way to the very last results (\ref{|G+-|<<1}) and (\ref{Temperature}). 

Thus, with the result (\ref{|G|>>1}) valid also for the case of quantized fields, we conclude that the observer would not detect any thermal spectrum of particles for small accelerations of his/her motion and/or large departures of the dispersion relation of such detected particles from the massless case. Similarly, from Eq.\,(\ref{|G+-|<<1}) we conclude that a thermal spectrum of particles would be detected for large accelerations of the observer and/or small departures of the dispersion relations of the detected particles from the massless case. Finally, with the result (\ref{Temperature}) we conclude that the detected spectrum of particles would look like a deformed Planck spectrum to which an apparent Unruh temperature can be associated. Such an apparent temperature is, in turn, different from the familiar Unruh temperature found for massless particles and can only be interpreted as an apparent temperature with a correction term that is frequency-dependent.

A more elaborate and pedagogical presentation of the fundamental difference between Fourier transforming only the phases that accompany the operators $a_k$ and $a^\dagger_k$ in the mode expansion of quantum fields and Fourier transforming the whole quantum field operator itself will be presented elsewhere. A link of the present approach with the Unruh-Dewitt detector method will then be presented there as well.

\section{Implications on the experimentally accessible classical analogue of the Unruh effect}\label{Implications}

Until very recently, no experimental observation of the Unruh effect was possible. The obvious reason being that the Unruh temperature, as given by Eq.\,(\ref{Temperature}) after setting in the latter $\omega=k$ and restoring to it the fundamental constants, becomes $T=\hbar a/(2\pi ck_B)$. Such an expression implies that even an acceleration as large as $10^{20}\,$m/s$^2$ would only produce a temperature which is even smaller than the $2.7\,$K of the ambient cosmic microwave background filling the Universe. Therefore, if one is not even able to detect such a thermal effect with experimentally accessible accelerations one can never hope for being able to test the deviations from thermality we derived here.

Fortunately, there have been many proposals in the literature to get around the technological limitations imposed on any attempt to observe the Unruh effect by focusing, instead, on attempts to observe a classical analogue of the effect \cite{SuperfluidAnalog,Retzker,Iorio,Rodriguez,Analog2018,Analog2020}. Accessible laboratory accelerations of the observer/detector are indeed sufficient for the effect to arise in this case. As it is already well known \cite{Classical1980,Takagi86,Classical1993,ClassicalRoots1,Classical2020}, the Unruh effect (and its deviation from thermality as we have seen in Sec.\,\ref{sec:II}) are not limited to the quantum fields of the Minkowski vacuum.  The effect, and the deviations therefrom we derived here, extend to classical waves as well. All we would need then to test experimentally the results we derived here is any kind of waves with a specific dispersion relation that departs from linearity. The setup proposed and realized in Ref.\,\cite{Analog2018} is, in this regard, what would best suit our needs.

The principle behind such a setup is to replace the vacuum fluctuations of Minkowski spacetime by the gravity waves on the surface of water subject to white noise. A laser beam is to be emitted perpendicularly toward the surface of the water to detect the ripples that play the role of vacuum fluctuations \cite{Analog2018}. By making such a laser beam translate horizontally with a constant acceleration, the laser spot traveling along the water surface, together with a camera recording the height of the illuminated spot of the water surface, would play the role of a Rindler observer (see Ref.\,\cite{Analog2018} for a description of the actual experiment). What makes such an experimental setup ideal for our present investigation is that one can easily modify the dispersion relation of the waves simply by adjusting the depth of the water in the container. For a gravity wave of wavelength $\lambda$ in shallow water of depth $h$, such that $h<0.05\lambda$, the dispersion relation of the wave becomes linear and takes the form $\omega(k)=k\sqrt{\varg h}$, where $\varg$ is the gravitational acceleration at the location of the experiment \cite{Dingemans}. However, for deep water, such that $h>\lambda/2$, the dispersion relation of the wave is nonlinear and it takes the form $\omega(k)=\sqrt{\varg k}$. The waves' phase velocity $c_p$ and group velocity $c_g$ in the case of deep water are then given by $c_p=\sqrt{\varg/k}$ and $c_g=c_p/2$; whereas for the case of shallow water both velocities become identical and reduce to the constant $\sqrt{\varg h}$. Therefore, to achieve an analogue of a Rindler observer in deep water, we only need to take $c_p$ to be the limiting speed in lieu of the speed of light $c$ in vacuum. Given that the experiment is performed with standing waves, the smallest wavenumber is $k=\pi/L$, where $L$ is the size of the container \cite{Analog2018}. Therefore, the limiting speed of the laser beam will be taken here to be $c=\sqrt{\varg L/\pi}$.    

Let us denote by $A(x,t)$ the amplitude of the ripples traveling on the water surface along the $x$-axis. We shall now adapt to our case the analysis presented in Ref.\,\cite{Analog2018} for the case of a linear dispersion relation of the water waves. However, before introducing noise into the water ripples as done in Ref.\,\cite{Analog2018}, let us first consider a pure monochromatic plane wave traveling on the water surface with a wavenumber $k$. For such a vibration mode, the boundary conditions imposed by the water container give rise to a standing wave that we can write in the form \cite{Analog2018},
\begin{equation}\label{StandingWaves}
A_k=\frac{1}{\sqrt{\omega(k)}}\,e^{-i\omega(k)t}\sin{(kx)}.    
\end{equation}
In accordance with Ref.\,\cite{Analog2018}, we have normalized this standing wave so that $(A_{k1},A_{k2})=\delta(k_1-k_2)$, where $(A_{k1},A_{k2})=i\int \left(A_{k1}^*\partial_t A_{k2}-A_{k2}\partial_t A_{k1}^*\right){\rm d}x$ is the time-invariant scalar product for modes obeying the wave equation. Expressing the phase in the standing wave (\ref{StandingWaves}) in terms of the proper time $\tau$ of the observer using the expression (\ref{ObserverPhase}) we derived above, and then Fourier transforming the result with an arbitrary angular frequency $\Omega$ using the prescription (\ref{Transform}), we arrive at the following expression, 
\begin{align}\label{StandingWavesFouried}
&\braket{A_k(\Omega)A^*_k(\Omega)}=\frac{1}{4\omega(k)}\left|g_+(\Omega)-g_-(\Omega)\right|^2\nonumber\\
&=\frac{1}{4\omega(k)}\left[|g_+(\Omega)|^2+|g_-(\Omega)|^2-g_+^*(\Omega)g_-(\Omega)-g_+(\Omega)g_-^*(\Omega)\right].
\end{align}
Here, the functions $g_{\pm}(\Omega)$ are now given by 
\begin{equation}\label{WaterTransform}
g_{\pm}(\Omega)=\frac{c}{a}\int_{0}^{\infty} y^{\pm\nu-1}e^{\pm i\left(\xi y+\frac{\eta}{y}\right)}{\rm d}y,
\end{equation}
and the parameters $\nu$, $\xi$ and $\eta$ are given by
\begin{equation}\label{AnalogXiEta}
\nu=i\frac{\Omega c}{a},\qquad \xi=\frac{kc^2-\omega(k)c}{2a},\qquad\eta=\frac{kc^2+\omega(k)c}{2a}.
\end{equation}
For a later convenience, we have restored here the constant $c$ to our parameters. Note also the slight difference in forms. The exponential in integrals (\ref{WaterTransform}) involves the sum $\xi y+\frac{\eta}{y}$, whereas the exponential in integrals (\ref{Transform}) involves the difference $\xi y-\frac{\eta}{y}$. These differences are due to having assigned, in accordance with Ref.\,\cite{Analog2018}, the phase factor $e^{-i\omega t}$ instead of $e^{i\omega t}$. Also, the parameters $\xi$ and $\eta$ are here switched. Nevertheless, both parameters are still positive as can be seen by plugging into the parameter $\xi$ our expressions of $\omega(k)$ for the water waves. 

Integrals (\ref{WaterTransform}) are evaluated in Eqs.\,(\ref{CosineSineCombined1}) and (\ref{CosineSineCombined2}) of Appendix \ref{A}. It is already evident from those expressions that, as with expression (\ref{g+-}), the term $\Gamma(\nu)$ which would lead to $(e^{2\pi\Omega c/a}-1)$ in the denominator is missing. The result (\ref{StandingWavesFouried}) clearly shows that, even for a monochromatic plane wave on the water surface, there is no trace of a Planck spectrum for arbitrary accelerations $a$ and dispersion relations $\omega(k)$. Moreover, we see that now there are, in addition, the interference terms $g_+^*(\Omega)g_-(\Omega)$ and $g_+(\Omega)g_-^*(\Omega)$ due to the reflected wave from the boundary. This interference will further affect the detected spectrum. However, as we saw in Sec.\,\ref{sec:II}, one can still probe limiting cases in the search for traces of the Planck spectrum. 
Therefore, in analogy with Eq.\,(\ref{|G|<<1}) of Sec.\,\ref{sec:II}, we expect that the only way to recover the Planck spectrum would be to have $\xi\eta\ll1$, which is equivalent to a large acceleration $a$ and/or a small departure from linearity of the dispersion relation. The latter condition can be realised experimentally by adjusting the depth of the water in the container. For $\xi\eta\gg1$, however, we see from the expansions in Eq.\,(\ref{Hankel>>1}) that the Planck spectrum has no chance to emerge.

Given that it is easier to experimentally control the acceleration $a$ of the laser beam than to adjust the depth of the water and the degree of departure from linearity of $\omega(k)$, we shall consider here the case of large accelerations. For that to be achieved, we need and acceleration such that $a\gg\tfrac{1}{2}\sqrt{k^2c^4-\omega^2c^2}$. Given that the wavenumber is conditioned by $k=m\pi/L$ \cite{Analog2018}, for any positive integer $m$, we deduce that for the case of deep water the acceleration of the laser beam need only satisfy $a\gg \tfrac{1}{2}\varg\sqrt{m(m-1)}$. These are indeed easily achievable accelerations in the lab. In what follows, we shall avoid the degenerate case $m=1$ for which $\xi=0$ and $\omega=kc$.

Inserting now the result (\ref{CosineSineCombined1}) for the first integral in Eq.\,(\ref{WaterTransform}), and then using the expansion (\ref{Hankel1RealExpansion}) of $H_\nu^{(1)}(z)$ for $z\ll1$, we easily obtain the amplitude $g_+(\Omega)$ for $\xi\eta\ll1$. Similarly, inserting the result (\ref{CosineSineCombined2}) for the second integral in Eq.\,(\ref{WaterTransform}), and then using the expansion (\ref{Hankel2RealExpansion}) of $H_\nu^{(2)}(z)$ for $z\ll1$, we obtain the amplitude $g_-(\Omega)$ for $\xi\eta\ll1$. The result at the leading order in $\xi\eta$ is the following unified expression,
\begin{equation}\label{WaterG+-<<1}
g_{\pm}(\Omega)\approx\frac{c}{a}\left[e^{-\frac{i\pi\nu}{2}}\Gamma(\mp\nu)\,\eta^{\pm\nu}+e^{\frac{ i\pi\nu}{2}}\Gamma(\pm\nu)\,\xi^{\mp\nu}\right].
\end{equation}
This result allows us to compute the various terms inside the square brackets in Eq.\,(\ref{StandingWavesFouried}). We find,
\begin{align}\label{LargeAExpectation}
\braket{A_k(\Omega)A^*_k(\Omega)}&\approx\frac{2\pi c}{\omega\,a\Omega\left(e^{\frac{2\pi\Omega c}{a}}-1\right)}\times\nonumber\\
&\quad\left[e^{\frac{\pi\Omega c}{a}}\sin\left(\theta-\frac{\Omega c}{a}\ln\eta\right)-\sin\left(\theta-\frac{\Omega c}{a}\ln\xi\right)\right]^2.
\end{align}
This result is similar to expression (\ref{|G+-|<<1}), but with a correction term in which the parameters $\eta$ and $\xi$ are are now `disentangled', and an overall multiplying factor of $1/\omega$ emerges. This is due to the extra deformation of the Planck spectrum caused by the interference with the reflected wave. 

Similarly to what we did for expression (\ref{|G+-|<<1}), we can probe the low-frequency region of the spectrum. For large accelerations and low frequencies such that $\Omega c\ll a$, the correction term in Eq.\,(\ref{LargeAExpectation}) can be expanded in the ratio $\Omega c/a$. As we are not looking for any apparent Unruh temperature here, we shall keep only terms up to the second order in $\Omega c/a$ inside the correcting factor. We arrive at the following approximate expression,
\begin{equation}\label{WaterCorrection}
   \braket{A_k(\Omega)A^*_k(\Omega)}\approx\frac{2\pi\Omega}{a^3\left(e^{\frac{2\pi\Omega c}{a}}-1\right)}\frac{\varg L^2}{\sqrt{m}}.
\end{equation}
This correction depends on the probed frequency $\Omega$ and can easily be measured experimentally in the laboratory as the integer $m$ just counts the number of harmonics of the monochromatic wave traveling on the water surface.

We are now going to examine the case of water waves permeated with white noise and follow step by step the analysis conducted in Ref.\,\cite{Analog2018}. Therefore, we should now express the amplitude $A(x,t)$ as a superposition of modes $A_k$ with coefficients $\alpha_k$ encoding the noise in the ripples,
\begin{equation}\label{AExpansion}
A(x,t)=\int_0^\infty\left(\alpha_kA_k+\alpha_k^*A_k^*\right){\rm d}k.    
\end{equation}
The mode amplitudes $\alpha_k$ represent Gaussian noise of uniform strength $I$, with the following averages: $\braket{\alpha_k}=0$ and $\braket{\alpha_{k_1}\alpha^*_{k_2}}=\frac{I}{2}\delta(k_1-k_2)$ \cite{Analog2018}. This is specifically what guarantees the emergence of a delta function in the frequency for the detected Doppler shifted noise.
Now, inserting the expression (\ref{StandingWaves}) into the expansion (\ref{AExpansion}) and using the result (\ref{WaterTransform}), we extract the Fourier transformed mode expansion as follows,
\begin{equation}\label{FourierAExpansion}
\tilde{A}(\Omega)=\!\!\int_0^\infty\!\!\!\frac{{\rm d}k}{2i\sqrt{\omega(k)}}\left(\alpha_k [g_+(\Omega)\!-\!g_-(\Omega)]\!+\!\alpha_k^*[\tilde{g}_-(\Omega)-\tilde{g}_+(\Omega)]\right).
\end{equation}
Here, the amplitudes $\tilde{g}_{\pm}(\Omega)$ stand for the expressions (\ref{g+-}) derived in Sec.\,\ref{sec:II}. Of course, in those expressions, the parameter $\eta$ should be taken with its absolute value $|\eta|$ since for the water waves we are considering here, we have $kc^2>\omega(k)c$. 

The total detected noise can be found by computing $\braket{\tilde{A}(\Omega_1)\tilde{A}^*(\Omega_2)}$ \cite{Analog2018}. For the case of a linear dispersion relation, such a calculation yields a Planckian spectrum accompanied by $\delta(\Omega_1-\Omega_2)$. This is rendered possible for two reasons. The first is the linearity in the relation $\omega(k)=k\sqrt{\varg h}$. The second, is that the amplitudes $g_{\pm}(\Omega)$ yield simply a factor of $\sin[\theta-\frac{\Omega c}{a}\ln(kc^2/a)]$ \cite{Analog2018}. These two facts lead to the appearance of the following transformed $\alpha(\Omega)$ in the integral (\ref{FourierAExpansion}),
\begin{equation}\label{LinearAphaIntegral}
\alpha(\Omega)\sim\int_0^\infty\frac{\alpha_k\,{\rm d}k}{k}\sin\left[\theta-\frac{\Omega c}{a}\ln\left(\frac{kc^2}{a}\right)\right].
\end{equation}
This integrated mode amplitude has been shown in detail in Ref.\,\cite{Analog2018} to be a Gaussian as well, in the sense that one still has the averages $\braket{\alpha_\Omega}=0$ and $\braket{\alpha_{\Omega_1}\alpha^*_{\Omega_2}}=\frac{I}{2}\delta(\Omega_1-\Omega_2)$. 

In our case, however, what we have is not only a nonlinear dependence on $k$ in the denominator because of $\omega(k)$, but also a nonlinear dependence on $k$ inside the functions $g_{\pm}(\Omega)$ which do not yield a simple sine function of $\ln k$. Moreover, this holds even for the case of large accelerations and/or small deviations of the dispersion relation from linearity. 
We see this by plugging the expression (\ref{WaterG+-<<1}) inside the first term multiplying $\alpha_k$ in the integral (\ref{FourierAExpansion}). We get, 
\begin{equation}\label{NonLinearAphaIntegral}
\alpha(\Omega)\!\sim\!\!\int_0^\infty\!\frac{\alpha_k\,{\rm d}k}{\sqrt{k}}\left[e^{\frac{\pi\Omega c}{2a}}\sin\left(\theta\!-\!\tfrac{\Omega c}{a}\ln\eta\right)+e^{\frac{-\pi\Omega c}{2a}}\sin\left(\theta\!-\!\tfrac{\Omega c}{a}\ln\xi\right)\right].
\end{equation}
When recalling that inside the parameters $\eta$ and $\xi$ there hide nonlinear functions of $k$ as well, it is evident that even the Gaussian structure guaranteed in the linear case by the transformed $\alpha(\Omega)$ is here lost. 

We thus clearly see the effect of noise on the detected spectrum when dispersion relations are allowed. For the case of large accelerations and/or small deviations of the dispersion relation from linearity, each monochromatic mode leads to a slight deviation from the Planck spectrum as given in Eq.\,(\ref{LargeAExpectation}). But, when all possible modes are combined into a Gaussian noise, thermalization is simply destroyed. Neither the large-acceleration regime nor the small deviations from linearity could then help restore the thermal spectrum. 
\section{Summary}
We have adapted the relativistic Doppler shift derivation of the Unruh effect to the case of a classical plane wave as well as to the case of a quantized field when both obey a modified dispersion relation of the form $\omega=\omega(k)$. The larger the difference $\omega(k)-k$, the larger the deviation of the dispersion relation from linearity one witnesses. We saw that the resulting general power spectrum of the detected waves/particles does not display any Planck-like pattern. As a result, we had to take into account the acceleration of the observer and the degree of deviation from linearity of the dispersion relation by distinguishing the two different cases of (i) small accelerations and/or large deviations and (ii) large accelerations and/or small deviations. In addition, we found that the approach applies successfully to the classical case and to the quantum case alike.

Among the many advantages of the approach is that one easily gains an intuitive explanation for the disappearance of the Planck spectrum for small accelerations and/or large deviations of the dispersion relation from linearity. The reason is that the relativistic Doppler shift in that case gives rise to a smothered Planck spectrum due to the great deviation of the dispersion relation from linearity. In contrast, for large accelerations and/or small deviations from linearity of the dispersion relation, an asymptotic Planckian-like spectrum emerges, to which one might associate an apparent Unruh temperature. The intuitive reason being that, although not a purely Planckian spectrum, one can regard the latter as so provided that one accepts to assign to it a frequency-dependent temperature, that one might thus call an apparent Unruh temperature. 

Another advantage of this approach is that it is so flexible that it easily accommodates the use of modified Lorentz transformations. Indeed, as we discussed in Sec.\,\ref{sec:II} and as we saw in detail in Appendix \ref{B}, it is possible to combine within this approach waves/fields with modified dispersion relations together with modified Lorentz transformations. The result is again physically very transparent and very intuitive in that it shows clearly how the nonlinearity of the dispersion relations spoils the Planck spectrum.

Yet, another advantage of the approach, mathematical in nature, is twofold. First, it is clear that to use Wightman's functions one needs first to relate the dispersion relation of the field to the deformation of the two-point functions of the time and space parameters $\Delta\tau$ and $\Delta x$, respectively \cite{InvariantPlanck}. Second, as we argued in the Introduction, the Bogoliubov transformations approach cannot be relied on either as (i) it heavily depends on the equations of motion of the deformed field and (ii) it only leads to the thermalization theorem which is already incapable of distinguishing even the massless case from the massive case \cite{Takagi86}.

Finally, as a prospect for an experimental test of our results, we have examined in detail the possibility of using gravity waves as an analogue substitute for the vacuum fluctuations of Minkowski spacetime. The role of the detector would be played by a light spot made by a laser beam emitted downwards perpendicularly towards the water surface. We examined two cases, the case of a pure monochromatic standing wave and the case of standing waves permeated with white noise. In the first case, thermality emerges corrected for large accelerations of the detector. In the second case, thermality is destroyed no matter what acceleration the detector has as long as the dispersion relation deviates from linearity. Our analysis thus showed great promise for experimentally testing the deviations from thermality we predicted here for waves with a dispersion relation. As we argued in Sec.\,\ref{Implications}, that goal is easily achievable when using waves on a water surface, for both the acceleration of the detector and the dispersion relation of such waves are easily accessible and easily adjustable experimentally.
\appendix
\section{Evaluating integrals (\ref{Transform}) and (\ref{StandingWavesFouried}) and finding their various expansions}\label{A}
In this appendix we gather the main identities that were useful in the text and we give the detailed calculations leading to the various formulas found in the text. 

We start by displaying the two useful integrals involving a power function and a trigonometric function \cite{FormulasBook},
\begin{align}\label{CosineSine}
\int_0^\infty y^{\nu-1}\cos\left(\xi y-\frac{\eta}{y}\right){\rm d}y&=2\left(\frac{\eta}{\xi}\right)^{\frac{\nu}{2}}K_\nu\left(2\!\sqrt{\xi\eta}\right)\cos\frac{\pi\nu}{2},\nonumber\\
\int_0^\infty y^{\nu-1}\sin\left(\xi y-\frac{\eta}{y}\right){\rm d}y&=2\left(\frac{\eta}{\xi}\right)^{\frac{\nu}{2}}K_\nu\left(2\!\sqrt{\xi\eta}\right)\sin\frac{\pi\nu}{2}.
\end{align}
In these integrals, both parameters $\xi$ and $\eta$ are assumed to be positive, which is the case in Sec.\,\ref{sec:II} where we deal with classical and quantum fields. Multiplying the second line by $i$ and adding it to the first, yields the result (\ref{g+-}) displayed in Sec.\,\ref{sec:II}.
Next, we have the following relation between the modified Bessel function $K_\nu(z)$ and Hankel's function $H_\nu^{(1)}(z)$ (valid for a complex number $z$ such that $-\pi<\arg z\leq\pi/2$):
\begin{equation}\label{BesselHankel}
K_\nu(z)=\frac{i\pi}{2}e^{\frac{i\pi\nu}{2}}H^{(1)}_{\nu}(iz).
\end{equation}

On the other hand, the useful series expansions for the functions $K_\nu(z)$ and $H_\nu^{(1)}(z)$ are given as follows. For large-magnitude arguments, $|z|\gg1$, we use the series expansion for the Bessel function \cite{FormulasBook} and then terminate the series at the zeroth order in $z$ as follows,
\begin{align}\label{K>>1}
K_\nu(z)&=\sqrt{\frac{\pi}{2z}}e^{-z}\left[\sum_{m=0}^{n-1}\frac{(2z)^{-m}\,\Gamma\left(\nu+m+\frac{1}{2}\right)}{m!\Gamma\left(\nu-m+\frac{1}{2}\right)}+\mathcal{O}\left(z^{-n}\right)\right]\nonumber\\
&\approx\sqrt{\frac{\pi}{2z}}e^{-z}.
\end{align}
For small-magnitude arguments, $|z|\ll1$, we find the series expansion for Hankel's function $H_\nu^{(1)}(z)$ by combining the series expansions of Bessel's first and second kind functions $J_\nu(z)$ and $Y_\nu(z)$, respectively, and then using $H_\nu^{(1)}(z)=J_\nu(z)+iY_\nu(z)$ \cite{FormulasBook}. From such a combination, we easily deduce indeed the following infinite series,
\begin{equation}\label{HankelExpansion}
H_\nu^{(1)}(z)=\sum_{m=0}^\infty\frac{(-1)^{m}z^{2m+\nu}}{m!\,2^{2m+\nu}}\left[\frac{1+i\cot(\pi\nu)}{\Gamma\left(m+\nu+1\right)}
-\frac{i\left(\frac{2}{z}\right)^{2\nu}\csc(\pi\nu)}{\Gamma\left(m-\nu+1\right)}\right].
\end{equation}
For small-magnitude arguments, $|z|\ll1$, we may terminate this infinite series in $m$ at the zeroth order to arrive at the following approximation for $H_\nu^{(1)}(iz)$:
\begin{align}\label{Hankel<<1}
H_\nu^{(1)}(iz)&\approx\frac{e^{i\frac{\pi\nu}{2}}z^{\nu}}{2^\nu}\Bigg[\frac{1+i\cot(\pi\nu)}{\Gamma\left(1+\nu\right)}
-\frac{i\left(\frac{2}{iz}\right)^{2\nu}\csc(\pi\nu)}{\Gamma\left(1-\nu\right)}\Bigg]\nonumber\\
&\approx-\frac{ie^{-i\frac{\pi\nu}{2}}}{\pi}\left[\Gamma(-\nu)\left(\frac{z}{2}\right)^{\nu}
+\Gamma(\nu)\left(\frac{z}{2}\right)^{-\nu}\right].
\end{align}
In the second step we have used the trigonometric identities $\cos(ix)=\cosh x$ and $-i\sin(ix)=\sinh x$ which hold for any real number $x$. In addition, we have also used the following two properties of the gamma function for a complex argument $z$ and a purely imaginary argument $ix$, respectively \cite{FormulasBook}:
\begin{equation}\label{Gamma}
\Gamma(1+z)=z\Gamma(z),\qquad |\Gamma(ix)|^2=\frac{\pi}{x\sinh(\pi x)}.
\end{equation}

We need now to evaluate integrals similar to those in Eq.\,(\ref{CosineSine}), but which involve $\cos(\xi y+\eta/y)$ and $\sin(\xi y+\eta/y)$, respectively. Such integrals can also be evaluated using the table of integrals in Ref.\,\cite{FormulasBook} (see, p. 480):
\begin{align}\label{CosineSineNegativeEta}
&\int_0^\infty\!\!y^{\nu-1}\!\cos\left(\xi y\!+\!\frac{\eta}{y}\right){\rm d}y=\!-\pi\left(\frac{\eta}{\xi}\right)^{\!\frac{\nu}{2}}\!\left[J_\nu(z)\sin\tfrac{\pi\nu}{2}+Y_\nu(z)\cos\tfrac{\pi\nu}{2}\right],\nonumber\\
&\int_0^\infty\!\!y^{\nu-1}\!\sin\left(\xi y\!+\!\frac{\eta}{y}\right){\rm d}y=\!\pi\left(\frac{\eta}{\xi}\right)^{\!{\frac{\nu}{2}}}\!\left[J_\nu(z)\cos\tfrac{\pi\nu}{2}-Y_\nu(z)\sin\tfrac{\pi\nu}{2}\right].
\end{align}
Here, $z$ stands for $2\!\sqrt{\xi\eta}$ and the functions $J_\nu(z)$ and $Y_\nu(z)$ are, respectively, Bessel's first and second kind functions. Multiplying the second line in Eq.\,(\ref{CosineSineNegativeEta}) by $i$ and adding it to the first line, and then using $H_\nu^{(1)}(z)=J_\nu(z)+iY_\nu(z)$ and $H_\nu^{(2)}(z)=J_\nu(z)-iY_\nu(z)$ \cite{FormulasBook}, yields,
\begin{align}
\int_0^\infty y^{\nu-1} e^{i\left(\xi y+\frac{\eta}{y}\right)}{\rm d}y&=i\pi e^{i\frac{\pi\nu}{2}}\left(\frac{\eta}{\xi}\right)^{\frac{\nu}{2}} H_{\nu}^{(1)}\left(2\sqrt{\xi\eta}\right),\label{CosineSineCombined1}\\
\int_0^\infty y^{-\nu-1} e^{-i\left(\xi y+\frac{\eta}{y}\right)}{\rm d}y&=-i\pi e^{i\frac{\pi\nu}{2}}\left(\frac{\eta}{\xi}\right)^{-\frac{\nu}{2}} H_{-\nu}^{(2)}\left(2\sqrt{\xi\eta}\right)\label{CosineSineCombined2}.
\end{align}
As these integrals directly involve Hankel's functions $H^{(1)}_\nu(z)$ and $H^{(2)}_\nu(z)$ rather than Bessel's function $K_\nu(z)$, we need to find the series expansions of the former for both $|z|\ll1$ and $|z|\gg1$. Using the series expansions for $J_\nu(z)$ and $Y_\nu(z)$ as given in Ref.\,\cite{FormulasBook}, we find the following series expansions for $|z|\gg1$:
\begin{align}\label{Hankel>>1}
H^{(1)}_\nu(z)&=\sqrt{\frac{2}{\pi z}}e^{i(z-\frac{\pi\nu}{2}-\frac{\pi}{4})}\left[\sum_{m=0}^{n-1}\frac{i^{m}\,\Gamma\left(\nu+m+\frac{1}{2}\right)}{(2z)^mm!\Gamma\left(\nu-m+\frac{1}{2}\right)}+\mathcal{O}\left(z^{-n}\right)\right]\nonumber\\
&\approx\sqrt{\frac{2}{\pi z}}e^{i(z-\frac{\pi\nu}{2}-\frac{\pi}{4})},\nonumber\\
H^{(2)}_\nu(z)&=\sqrt{\frac{2}{\pi z}}e^{i(\frac{\pi\nu}{2}+\frac{\pi}{4}-z)}\left[\sum_{m=0}^{n-1}\frac{(-i)^{m}\,\Gamma\left(\nu+m+\frac{1}{2}\right)}{(2z)^mm!\Gamma\left(\nu-m+\frac{1}{2}\right)}+\mathcal{O}\left(z^{-n}\right)\right]\nonumber\\
&\approx\sqrt{\frac{2}{\pi z}}e^{i(\frac{\pi\nu}{2}+\frac{\pi}{4}-z)}.
\end{align}
For $|z|\ll1$, we already have the expansion (\ref{Hankel<<1}) for $H_\nu^{(1)}(iz)$ which for a real argument $z$, leads to 
\begin{equation}\label{Hankel1RealExpansion}
H_\nu^{(1)}(z)\approx-\frac{ie^{-\frac{i\pi\nu}{2}}}{\pi}\left[e^{\frac{-i\pi\nu}{2}}\Gamma(-\nu)\left(\frac{z}{2}\right)^{\nu}
+e^{\frac{i\pi\nu}{2}}\Gamma(\nu)\left(\frac{z}{2}\right)^{-\nu}\right].
\end{equation}
For $H_\nu^{(2)}(z)$, we find the following expansion when $z\ll1$: 
\begin{align}\label{Hankel2RealExpansion}
H_\nu^{(2)}(z)&=\sum_{m=0}^\infty\frac{(-1)^{m}z^{2m+\nu}}{m!\,2^{2m+\nu}}\left[\frac{1-i\cot(\pi\nu)}{\Gamma\left(m+\nu+1\right)}
+\frac{i\left(\frac{2}{z}\right)^{2\nu}\csc(\pi\nu)}{\Gamma\left(m-\nu+1\right)}\right]\nonumber\\
&\approx\frac{ie^{\frac{i\pi\nu}{2}}}{\pi}\left[e^{\frac{i\pi\nu}{2}}\Gamma(-\nu)\left(\frac{z}{2}\right)^{\nu}
+e^{-\frac{i\pi\nu}{2}}\Gamma(\nu)\left(\frac{z}{2}\right)^{-\nu}\right].
\end{align}
\section{Working with deformed Lorentz transformations}\label{B}
In this appendix we examine the case of observers (detectors) using the waves/fields they are hit by as probes of their spacetime location, requiring, as a consequence, the use of deformed Lorentz transformations when computing the perceived phase $\phi(\tau)$. Now, given the multitude of proposals introduced in the literature for modified dispersion relations and their corresponding modified Lorentz transformations, we shall not consider here every single model introduced in the literature, but focus instead on the very general model reported in Ref.\,\cite{DSRInx2007}.

For a particle of angular frequency $\omega(k)$ and wave number $k$ relative to an inertial frame, the modified Lorentz transformation for time in $1+1$ dimensions is given by \cite{DSRInx2007},
\begin{equation}\label{MLt}
t'=\frac{f(p)}{f(p')}\left(\frac{t-\frac{kk'}{\omega\omega'}\varv x}{\sqrt{1-\frac{k'^2}{\omega'^2}\varv^2}}\right).
\end{equation}
Here, $p$ is the momentum of the particle in the original frame and $f(p)$ is an arbitrary function that relates the angular frequency to the energy of the particle: $\omega=Ef(p)$. The transformed quantities $p'$, $\omega'$ and $k'$ represent the properties of the particle in a frame moving with instantaneous velocity $\varv(t)$. From this expression we extract the relation between the element of the proper time ${\rm d}\tau$ in terms of ${\rm d}t$ as follows:
\begin{align}\label{tau}
{\rm d}\tau=\frac{f(p')}{f(p)}\sqrt{1-\frac{k'^2\varv^2}{\omega'^2}}\,{\rm d}t.
\end{align}
Next, by using the transformation of the position \cite{DSRInx2007},
\begin{equation}\label{MLx}
x'=\frac{f(p)}{f(p')}\left(\frac{x-\frac{\omega k'}{k\omega'}\varv t}{\sqrt{1-\frac{k'^2}{\omega'^2}\varv^2}}\right),
\end{equation}
we easily find the formula for the transformation of velocities, from which, in turn, we deduce the relation between the acceleration ${\rm d}\varv/{\rm d}t$ in the rest frame of the laboratory and the constant proper acceleration $a$ in the instantaneous rest frame of the observer. The result is,
\begin{equation}\label{Propera}
a=\frac{f(p)}{f(p')}\frac{\omega k'}{k\omega'}\frac{{\rm d}\varv/{\rm d}t}{\left(1-\frac{k'^2}{\omega'^2}\varv^2\right)^{3/2}}.
\end{equation}
By integrating this equation after taking the initial condition $\varv=0$ at $t=0$, we extract the velocity $\varv$ in terms of the proper acceleration $a$ and the time $t$ as follows:
\begin{equation}\label{v(t)}
\varv(t)=\frac{f(p')}{f(p)}\frac{k\omega'}{\omega k'}\frac{at}{\sqrt{1+\frac{f^2(p')}{f^2(p)}\frac{k^2}{\omega^2}a^2t^2}}.
\end{equation}
Substituting this result into Eq.\,(\ref{tau}) and integrating, after taking the initial condition $\tau=0$ at $t=0$, allows us to find the time $t$ in terms of the proper time $\tau$ and then the position $x$ in terms of the proper time $\tau$ as well. We find,
\begin{equation}\label{MLx,t}
t(\tau)=\frac{\omega f(p)}{kf(p')}\frac{\sinh{(ak\tau/\omega)}}{a},\quad x(\tau)=\frac{\omega\omega'f(p)}{kk'f(p')}\frac{\cosh{(ak\tau/\omega)}}{a}.
\end{equation}
Finally, substituting these expressions of $t(\tau)$ and $x(\tau)$ inside the phase $e^{i\phi_{\pm}(t,x)}=e^{i[\omega(k) t\pm kx]}$ of Minkowski spacetime, we find the effective phase detected by the observer to be of the form,
\begin{equation}\label{MLObserverPhase}
e^{i\phi_{\pm}(\tau)}=\exp\left(\frac{if(p)\omega}{f(p')}\left[\frac{\omega k'\pm\omega'k}{2akk'} e^{ak\tau/\omega}-\frac{\omega k'\mp\omega'k}{2akk'}e^{-ak\tau/\omega}\right]\right).
\end{equation}
Let us keep in mind here that both angular frequencies $\omega(k)$ and $\omega'(k')$ in this expression depend nonlinearly on their respective wave numbers $k$ and $k'$ in the original and the new frames, respectively. 

To get the shape of the spectrum as detected by the accelerating observer, we Fourier transform the $\tau$-dependent phase (\ref{MLObserverPhase}) using an arbitrarily chosen angular frequency $\Omega$ among the continuous spectrum of frequencies available to the observer. Once again, as the transform consists in evaluating the integral, $g_{\pm}(\Omega)=\int_{-\infty}^{+\infty} e^{i\Omega\tau}e^{i\phi_{\pm}(\tau)}{\rm d}\tau$, we are going to perform the change of variable $e^{ak\tau/\omega}=y$. The Fourier transform then takes the form, 
\begin{equation}\label{MLTransform}
g_{\pm}(\Omega)=\frac{1}{a}\int_{0}^{\infty} y^{\pm\nu-1}e^{\pm i\left(\xi y-\frac{\eta}{y}\right)}{\rm d}y.
\end{equation}
Here, we have distinguished the spectrum $g^{+}(\Omega)$ of the left-moving modes from the spectrum $g^{-}(\Omega)$ of the right-moving ones. We have also set, for convenience,
\begin{equation}\label{MLXiEta}
\nu=i\frac{\omega\Omega}{ak},\;\quad \xi=\frac{\omega f(p)}{f(p')}\frac{\omega k'+\omega'k}{2akk'},\;\quad\eta=\frac{\omega f(p)}{f(p')}\frac{\omega k'-\omega'k}{2akk'}.
\end{equation}
To evaluate integral (\ref{MLTransform}) we follow the same steps described in Section \ref{sec:III}. As this integral is identical to integral (\ref{Transform}), we find again that the amplitudes are given in terms of the modified Bessel function $K_\nu(z)$ by,
\begin{equation}\label{DRSg+-}
g_{\pm}(\Omega)=\frac{2\,e^{i\frac{\pi\nu}{2}}}{a}\left(\frac{\eta}{\xi}\right)^{\pm\frac{\nu}{2}} K_{\pm\nu}\left(2\sqrt{\xi\eta}\right).
\end{equation}
The important difference, however, is that now the exponent $\nu$ that is responsible for giving rise to the Planck spectrum via the crucial term $e^{-\pi\Omega/2a}$ (emerging from the factor $e^{i\pi\nu/2}$ in this expression) is here replaced by the term $e^{-\pi\omega\Omega/2ak}$. The detected angular frequency $\Omega$ is thus never isolated from the waves' phase velocity $\omega(k)/k$. As a consequence, the Planck spectrum can never be ``purified'' from the effect of the nonlinear dispersion relation of the detected waves ---no matter how large the acceleration $a$ is or how close to linearity the dispersion relation is--- as long as the latter is not exactly linear as in the massless case.

\section*{Acknowledgments}
The authors are grateful to the anonymous referee for the pertinent comments and insightful remarks that improved our manuscript. This work is supported by the Natural Sciences and Engineering Research Council of Canada (NSERC) Discovery Grant (No. RGPIN-2017-05388).


\end{document}